# MINFLUX achieves molecular resolution with minimal photons

Lukas Scheiderer[1,2,3]*, Zach Marin[1,2]*, Jonas Ries[1,2,3,4]

[1] Max Perutz Labs, Vienna Biocenter Campus (VBC), Vienna, Austria
[2] University of Vienna, Center for Molecular Biology, Department of Structural and Computational Biology, Vienna, Austria
[3] European Molecular Biology Laboratory, Cell Biology and Biophysics, Heidelberg, Germany
[4] University of Vienna, Faculty of Physics, Vienna, Austria

* Equal contribution
Correspondence: Jonas.ries@univie.ac.at

## Abstract

Optical super-resolution microscopy is a key technology for structural biology that offers high imaging contrast and live-cell compatibility. Minimal (fluorescence) photons flux microscopy, or MINFLUX, is an emerging super-resolution technique that localizes single fluorophores with high spatiotemporal precision by targeted scanning of a patterned excitation beam featuring a minimum. MINFLUX offers super-resolution imaging with nanometer resolution. When tracking single fluorophores, MINFLUX can achieve nanometer spatial and sub-millisecond temporal resolution over long tracks, greatly outperforming camera-based techniques. In this review, we present the basic working principle of MINFLUX and explain how it can reach high photon efficiencies. We then outline the advantages and limitations of MINFLUX, describe recent extensions and variations of MINFLUX and, finally, provide an outlook for future developments.

## Introduction

Continuous advances in electron microscopy (EM) over the last century now make it possible to resolve structures of proteins and their complexes with angstrom resolution even in their native cellular context[1] (Figure 1a). However, to reach a resolution that is sufficient to identify individual proteins in the electron densities, typically many identical structures need to be averaged. Optical super-resolution microscopy, on the other hand, uses fluorescent labels that allow investigation of individual structures without averaging and with high contrast. Unlike EM, it is applicable to living cells and can directly measure dynamics. In the last two decades, its resolution has been pushed to the molecular scale. Thus, it is becoming an increasingly important complementary technology for structural cell biology[2].

Fluorescence microscopes are among the most important technologies for biology. They either use camera-based detection (widefield microscope) or a scanned laser focus and a point detector in combination with a pinhole for rejecting out-of-focus fluorescence (confocal microscope, Figure 1b). Their spatial resolution, however, is limited to >200 nm, about half the wavelength $\lambda$ of the excitation light used and far worse than what is useful for structural biology. The size (standard deviation of the Gaussian approximation, corresponding to the microscope resolution/2.355) of the point spread function (PSF) depends on the numerical aperture of the objective and is[3]

$$\sigma_{\text{PSF}} = \frac{0.21\lambda}{NA} \tag{1}$$

Super-resolution microscopy (SRM) technologies use smart tricks to circumvent this resolution limit for diffraction-unlimited resolution. Stimulated emission depletion (STED) microscopy[4,5] narrows the size of the excitation beam of a confocal microscope by super-positioning of a donut-shaped depletion beam that quenches the fluorescence emission outside of the donut center (Figure 1c). This usually leads to an up to ~10-fold improved spatial resolution, which is determined by the intensity $I$ of the STED laser compared to the saturation intensity $I_S$, leading to a PSF size of:

$$\sigma_{\text{STED}} = \frac{\sigma_{\text{PSF}}}{\sqrt{1 + I/I_S}}. \qquad (2)$$

Single molecule localization microscopy[6,7] (SMLM) relies on labeling target structures with fluorophores that can be switched between a long-lived dark state and a bright state. On-switching of only a small subset of fluorophores in each of many camera frames leads to sparse images of single fluorophores. Their positions can then be determined with nanometer precision by fitting a model of the PSF to the recorded pixel intensities (Figure 1d). Because of shot noise, the intensity in each pixel is a random variable from a Poisson distribution, and the localization precision depends strongly on the number of detected fluorescence photons $N$:

$$\sigma_{\text{SM}} = \frac{\sigma_{\text{PSF}}}{\sqrt{N}} \qquad (3)$$

Under realistic conditions, any fluorescence background degrades $\sigma_{\text{SM}}$, which can be calculated, e.g., as described in Mortensen et al.[8]. The use of fluorophores that emit a large number of photons $N$ leads to a localization precision of a few nanometers[9], and averaging over repeated localizations of the same target protein has pushed the localization precision towards the angstrom regime[10]. In contrast to $\sigma_{\text{PSF}}$ and $\sigma_{\text{STED}}$, which denote the physical size of the PSF, the localization precision $\sigma_{\text{SM}}$ denotes the expectation value of the standard deviation of the fluorophore position.

In SMLM, a super-resolution image is reconstructed by combining millions of fluorophore positions imaged in thousands of camera frames. The principle of precise localization of single emitters with a camera is also key in widefield single-fluorophore tracking techniques[11], which quantify the spatial dynamics of single emitters.

MINFLUX[12], developed 2017 by the group of Stefan Hell, outperforms the SMLM localization precision limit (Eq. 3) by targeted probing of a single fluorophore with a patterned beam that features an intensity minimum (Figure 1e), reaching unprecedented photon-efficiencies. In combination with switchable fluorophores, it allows for SMLM imaging with single-nanometer precision with few photons. For tracking, it can improve the spatial resolution, temporal resolution and track length each by around one order of magnitude compared to camera-based techniques.

## Principle of MINFLUX

How does MINFLUX achieve a better localization precision than possible with camera-based or confocal detection?

Let us consider a confocal microscope image of a single fluorophore displaying the (close-to-Gaussian) PSF of the microscope. PSF fitting of this image would lead to the same localization precision as with a camera (Eq. 3). Interestingly, not all pixels in this image carry equal position information: pixels at the flanks of the Gaussian show a strong intensity dependence with small variations of the fluorophore position, whereas pixels at the flat PSF center carry little position information. If we had prior information on the approximate position of the fluorophore, we could omit the bright, low-information pixels and achieve a similar localization precision with fewer photons from the flanks.

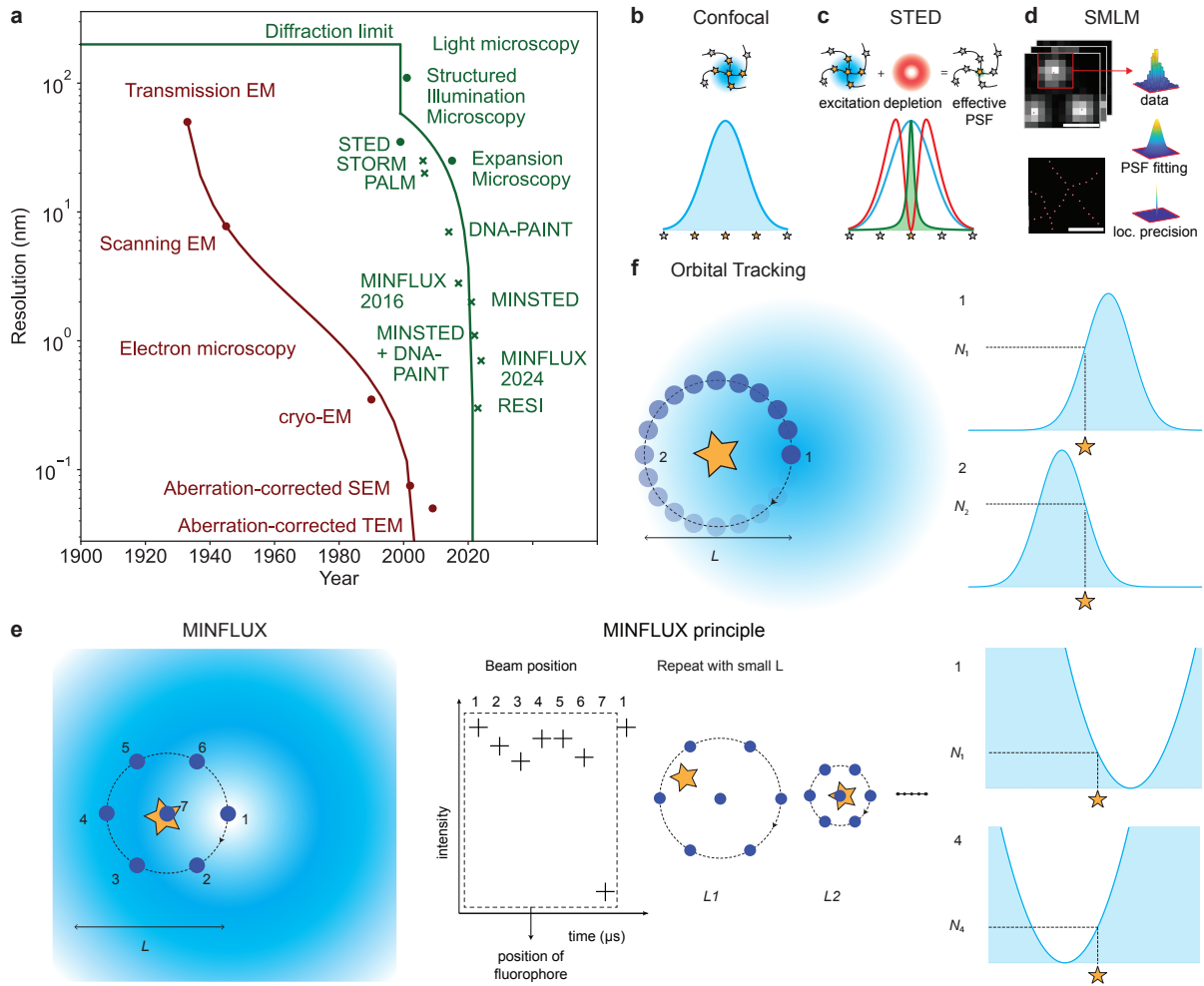

**Figure 1. Comparison of super-resolution fluorescence microscopy (SRFM) techniques. a,** SRFM has rapidly advanced in the last 20 years[4–7,9,10,12,19,20,34,58–61], approaching the resolution of electron microscopy[62–67] with the advantages of molecular specificity and live-cell compatibility. Single-molecule localization methods (SMLM), including photoactivated localization microscopy (PALM)[6], stochastic optical reconstruction microscopy (STORM)[7,59], DNA point accumulation in nanoscale topography (DNA-PAINT)[9], MINSTED[19] and MINFLUX[12,34], are marked with an ×. **b, Confocal microscopy** illuminates all fluorophores within the excitation PSF and achieves a resolution of ~200 nm at best. **c, STED microscopy** combines confocal excitation (blue) with a donut-shaped depletion beam (red) to quench fluorescence in the periphery, which ensures only signal from fluorophores at the low-intensity center (green) are detected, achieving a resolution of ~35 nm. **d, SMLM** uses blinking fluorophores covering the entire field of view, where few fluorophores emit in any single camera frame. Each fluorophore position is determined by fitting to a PSF model. An image with ~5 nm precision is reconstructed from the positions collected across all frames. **e, Orbital Tracking** scans a Gaussian beam circularly around a fluorescent particle to improve particle localization precision by up to ~1.7x compared to SMLM. The particle position is inferred from measured intensities, which are highly dependent on fluorophore position near the steep portion of the Gaussian PSF. The scan path is continuously re-centered around the fluorophore. **f, MINFLUX** uses a donut-shaped PSF with a near-zero intensity center that is sequentially placed at several positions around the fluorophore. From the measured intensities, the fluorophore position is estimated. Iterative MINFLUX combines several localization steps with decreasing diameter of the scan path $L$ to achieve single-nanometer localization precisions. A steep PSF gradient combined with low intensities near the donut minimum improves the localization precision per photon around fivefold compared to SMLM.

To investigate the precision with which we can determine the position of a fluorophore by targeted probing with focused light, we will consider a simple 1D scheme without background and with a PSF of shape $f(x_0 - x_i)$ positioned at $x_i = 0$. For a fluorophore at position $x_0$, we detect $N = I_0 f(x_0)$ photons. From such a single measurement we can retrieve the fluorophore position as $x_0 = f^{-1}(N/I_0)$, provided we know its brightness $I_0$. In a linear approximation, the error of the position $\delta x$ is proportional to the error $\delta N$ in measuring the photons $N$ (Figure 2a):

$$\delta N = I_0 \, f'(x_0) \, \delta x_0 \qquad (4)$$

Here, $f'(x_0) = \partial f/\partial x|_{x_0}$ is the derivative of $f$ along $x$, i.e., the gradient of the PSF, evaluated at $x_0$. Because the detection of photons is a Poisson process, $\delta N = \sqrt{N}$. We can substitute $I_0 = N/f(x_0)$ and solve for $\delta x_0$:

$$\delta x_0 = \left| \frac{1}{\sqrt{N}} \frac{f(x_0)}{f'(x_0)} \right| \qquad (5)$$

where we have taken the absolute value since localization precision is always positive. Thus, in a first approximation, the position error scales inversely with $\sqrt{N}$ and otherwise depends only on the shape of the PSF: both a low magnitude $f(x)$ and a strong gradient $f'(x)$ at the emitter position $x_0$ lead to a good localization precision. For realistic measurements, the fluorophore brightness is typically not known, and we need to perform additional measurements.

The idea of using a strong gradient to localize an emitter is realized in orbital tracking (Figure 1f), where the steep part of the PSF is scanned in a circular path around an assumed position of a single fluorescent particle (to our knowledge, tracking of targets labeled with single fluorophore molecules has not been demonstrated with this technique)[13–15]. Key for this approach is to iteratively estimate the position of the particle from the intensity variation during the orbital scan and to recenter the midpoint of the circular beam path on the best position estimate. Using a 2D Gaussian approximation of the PSF:

$$I_{\max}(\bar{x}_0 - \bar{x}) = I_0 e^{-\frac{(\bar{x}_0 - \bar{x}_i)^2}{2\sigma_{\text{PSF}}^2}} \qquad (6)$$

with $I_{\max}(\bar{x}_0 - \bar{x}_i)$ denoting the intensity measured at position $\bar{x}_i = (x_i, y_i)$, $I_0$ the amplitude of the excitation beam, and $\bar{x}_0$ the position of the emitter, we can calculate the localization precision for a fluorophore that is centered at a circular orbit with a radius $L/2$, in absence of any background fluorescence, as[16]:

$$\sigma_{\text{OT}} = \sigma_{\text{SM}} \frac{\sqrt{8} \, \sigma_{\text{PSF}}}{L} \qquad (7)$$

Thus, we achieve an improvement over diffraction-limited PSF localization if we choose the diameter of the orbit $L > \sqrt{8} \, \sigma_{PSF}$, which is in agreement with the expected value for orbital radius in the original orbital tracking paper, and results in probing the fluorophore at the half maximum of the PSF[13]. In practice, background fluorescence causes a drop in signal-to-background ratio as the Gaussian beam moves away from the fluorophore and a suitable choice is $L \approx 5 \, \sigma_{\text{PSF}}$, which results in probing the fluorophore near, but before, where the Gaussian tail flattens[17], leading to an improvement of the localization precision by a factor of $\sigma_{\text{OT}} \approx 0.6 \, \sigma_{\text{SM}}$.

To improve the localization precision even further, we can generate PSFs with steeper intensity gradients, like the narrow and steep STED PSF. When used for orbital tracking[18], this approach is termed MINSTED[19,20]. An alternative approach is to use a PSF featuring a local intensity minimum with (near-)zero intensity. Here, fluorophores experience a substantial change in excitation intensity with the distance from the minimum, while the intensity at the fluorophore position is still low (Figure 1e), leading to a high position information per detected photon. This is the principle of MINFLUX.

During a MINFLUX measurement, the minimum of a patterned excitation beam, typically a donut beam (not to be confused with the commonly used donut-shaped beam in STED used for depletion instead of excitation) is positioned at different locations around the assumed fluorophore position (Figure 1e). From the detected number of photons at each location and from the precisely known position of the beam, an improved position estimate can be calculated. In case we have a fluorophore centered in the field of view (FOV), a perfect intensity minimum, and no background, the 2D localization precision is[12]

$$\sigma_{\mathrm{MF}} = \frac{L}{\sqrt{8N}} \qquad (8)$$

Here, $L$ is the diameter of the scan pattern, assumed smaller than the size of the PSF, and $N$ the number of detected photons. Compared to the diffraction-limited localization precision (Eq. 3), we gain in two ways: we find an additional factor of $8^{-1/2}$ and the size of the scan pattern $L$ can be chosen much smaller than the PSF size $\sigma_{\mathrm{PSF}}$.

To intuitively understand why MINFLUX has a high localization precision for a low number of detected photons, let us assume that we already know the position of the fluorophore and can place the dark donut center precisely over it. As the fluorophore does not see any light, we will not detect any photons. But if we move it ever so slightly, it will see some light and emit some photons, and we can easily distinguish no photons from a few photons and from more photons. This is where the high position sensitivity per photon comes from.

So why not simply use an infinitely small $L$ for infinite resolution? The first reason is that in realistic measurements the intensity minimum is not truly zero because of PSF imperfections and fluorescence background. The result is higher shot noise and consequently a worse localization precision, as discussed in detail below. The second reason is that Eq. 8 is only valid if the fluorophore is close to the center of the scan pattern, typically within $L$. But in the beginning of the experiment, only an approximate position of the fluorophore is known and a large $L$ needs to be chosen. This limitation is overcome with iterative MINFLUX[21], which zooms in on the emitter by decreasing $L$ in each iteration while re-centering the scan pattern on the newly estimated position, allowing for a small final $L$ with high localization precision (Figure 1e). Note that all photons detected during all iterations are part of the total photon budget and should be included in estimations of the photon efficiency.

Let us use the simple 1D localization precision described in Eq. 5 to examine the difference between measuring a fluorophore position with an intensity maximum (Figure 2a), as in scanning confocal and orbital tracking, or an intensity minimum (Figure 2b), as in MINFLUX. For a fixed PSF width determined by the optical system, the maximum exhibits a larger photon count and therefore a larger uncertainty $\delta N = \sqrt{N}$ than the minimum at the fluorophore position $x_0$. Furthermore, the gradient of the minimum can be stronger than that of the maximum at this point. We can investigate this analytically by approximating the minimum with a quadratic function

$$I_{\min}(x_0 - x_i) = I_0 \left( \frac{(x_0 - x_i)^2}{2\,\sigma_{\mathrm{q}}^2} + b \right). \qquad (9)$$

Here, $I_{\min}(x_0 - x_i)$ denotes the collected intensity, i.e. number of photons, at position $x_i$, $I_0$ is a proportionality factor that describes how the brightness of the emitter depends on the illumination laser power, $x_0$ is the position of the emitter, and $\sigma_{\mathrm{q}}$ parametrizes the steepness of the PSF. $b$ describes a background either due to an imperfect zero of the PSF or autofluorescence of the sample or out-of-focus fluorescence from nearby fluorophores. As all these background contributions scale with the amplitude of the excitation, we multiply $b$ with $I_0$ in Eq. 9, thus $b$ itself is independent of the excitation intensity. If we set the background term to zero, we can substitute the 1D case of Eq. 6 and Eq. 9 into Eq. 5 to see how $\delta x_0$ varies as a function of $x_0$. In this simple scheme,

$$\delta x_{0,\max} = \frac{\sigma_q^2}{x_0\sqrt{N}} \quad \text{and} \quad \delta x_{0,\min} = \frac{x_0}{2\sqrt{N}} \tag{10}$$

For the minimum case, the localization precision $\delta x_{0,\min}$ is improved the closer we measure to the center of the PSF, whereas $\delta x_{0,\max}$ improves the further away we measure.

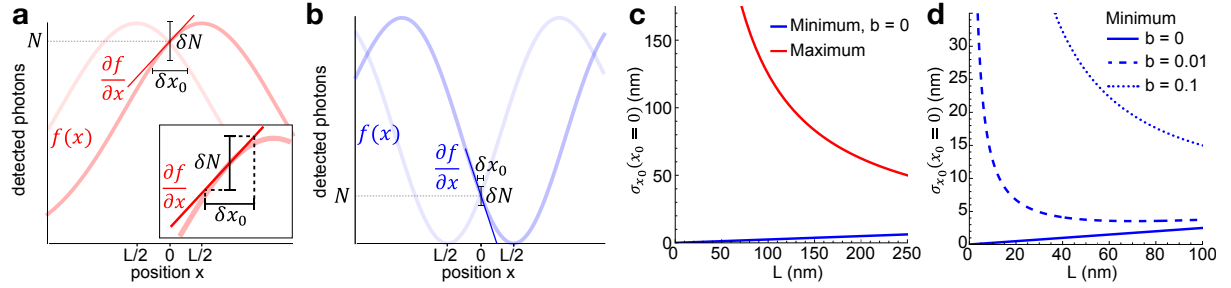

**Figure 2. Localization with maxima vs. minima in 1D**. **a,** Two maximum PSF measurements at $x_i = \{-L/2, L/2\}$ of an emitter at position $x_0 = 0$. Inset: The position error $\delta x_0$ can be calculated from the error $\delta N$ in measuring the photons $N$ if the gradient of the PSF $\partial f/\partial x$ is known. When the emitter is near the PSF center, $\delta N = \sqrt{N}$ is large and the gradient is small, resulting in a poor localization precision (LP) $\delta x_0$. **b,** Two minimum PSF measurements. When the emitter is near the PSF center, $\delta N$ is small, and the gradient is large, resulting in a good LP $\delta x_0$. **c,** LP (CRB) for illumination of a fluorophore with two minima vs. two maxima. Here, $x_0 = 0$, $b = 0$, $N_{\text{tot}} = 100$ photons and $\sigma_{\text{PSF}} = 250$ nm. **d,** Even with background, minima outperform maxima PSF measurements. For the minimum case, at $L = 100$, $b = 0.01$ corresponds to a signal-to-background ratio (SBR) of 3 and $b = 0.1$ corresponds to an SBR of 1.2.

The best achievable localization precision can be calculated as the Cramér-Rao (lower) bound (CRB) as detailed in Masullo et al[17]. Here we follow this approach and extract the localization precision $\sigma_{x_0}$ of the emitter position measurement from the corresponding matrix element. In a realistic experiment, the brightness of the fluorophore and with it the total number of detected photons, $N_{\text{tot}} = \sum N_i$, and the fluorophore position $x_o$ are unknown. Thus, we need at least 2 measurements to recover these two unknown parameters. If the background $b$ is not known, we need at least 3 measurements to determine $N_{\text{tot}}$, $x_0$ and $b$[16]. Let's assume the background is known, and we measure at $x_i = \{-L/2, L/2\}$. Then, we can calculate the localization precision of the minimum case (Eq. 9) as[16]

$$\sigma_{\min} = \frac{1}{4L\sqrt{N_{\text{tot}}}} \frac{(L^2 + 4x_0^2 + 8b\sigma_q^2)\sqrt{(L^2 - 4x_0^2)^2 + 16b\sigma_q^2(L^2 + 4x_0^2 + 4b\sigma_q^2)}}{L^2 - 4x_0^2 + 8b\sigma_q^2} \tag{11}$$

which reduces to the reported minimum 1D MINFLUX precision[12]

$$\sigma_{\text{MF,1D}} = \sigma_{\min}(x_0 = 0, b = 0) = \frac{L}{4\sqrt{N_{\text{tot}}}} \tag{12}$$

when $x_0 = 0$ and $b = 0$. Note that this is equivalent to the simple approximation, Eq. 10, evaluated at $x_0 = L/2$.

We could alternatively localize the same fluorophore with two Gaussian maxima (Eq. 6). This results in a precision similar to orbital tracking (Eq. 7). We can see (Figure 2c) that this performs poorly compared to the minimum, as suggested by the simple approximation, even when there is reasonably high background (Figure 2d).

To summarize, the conceptual requirements for MINFLUX to reach photon-efficient, diffraction-unlimited resolution are:

- Single isolated fluorophores.
- A PSF with a near-zero minimum to generate a steep intensity gradient.
- Targeted positioning of the PSF in a defined pattern.
- Iterative real-time feedback on the position of the scan pattern or sufficiently precise prior information on the fluorophore position.

In addition to these fundamental requirements, the success of a MINFLUX experiment depends on several practical requirements and considerations:

Attaching a fluorophore to the target biomolecule comes with two challenges. Firstly, the labeling process should not influence the function of the protein. Secondly, the distance of the fluorophore to the target protein should be minimal, otherwise this linkage error will be the main source of the position error rather than the photon limit on localization precision. The common approach of labeling with antibodies can lead to linkage errors of up to 15 nm. Nanobodies, self-labeling enzymes (e.g. SNAP-tag or HaloTag), unnatural amino acids or genetic tagging with fluorescent proteins have lower linkage errors and are preferable for MINFLUX[2].

Many fluorophores show intensity changes (flickering) on the micro- to millisecond time scale[22,23]. If, for example, the emitter goes to an off state when the minimum is at a peripheral position, fewer photon will be detected, leading to a wrong position calculation by the MINFLUX estimator[24], which assumes constant intensities. This error can be reduced by averaging over the flickering by using many fast scan cycles for a single localization. Alternatively, one can select more suitable fluorophores for MINFLUX with inherently low flickering and high brightness. Live-cell measurements require live-cell compatible fluorophores and come with the danger that phototoxicity induced by the MINFLUX laser can potentially perturb the process under investigation. In general, the laser intensities in MINFLUX ($\sim$10-50 kW/cm$^2$) are comparable to those used in confocal imaging[21] and hence are roughly an order of magnitude higher than for SMLM[25,26]. But, as the cell is illuminated only in a small region corresponding to the size of the PSF, the average laser intensities a cell sees can be much smaller than in widefield excitation. In any case, careful controls are necessary to exclude artifacts by photo toxicity in any live-cell fluorescence measurement.

At the nanoscale, the resolution is in many cases not only limited by the number of photons but by instabilities of the microscope (drift, vibrations), which can easily exceed the nominal localization precision, underscoring the need for extremely precise stabilization of the microscope.

As the signal of the target fluorophore comes from the vicinity of the dark donut center, which is easily outshone by signal generated by perturbing fluorophores illuminated by the bright part of the PSF, the risk of multiple emitters being in their on-state needs to be reduced by activating only a very low fraction of emitters at a time. This is why MINFLUX is implemented with confocal detection, where the pinhole rejects out-of-focus and peripheral fluorescence. For highly dense samples with significant (auto-)fluorescent background, MINSTED might be superior to MINFLUX due to its smaller PSF size[27].

## MINFLUX imaging

MINFLUX requires low fluorophore densities and low fluorescence background. As such, to use MINFLUX for super-resolution imaging, it needs to be combined with SMLM, i.e.

blinking/photoactivatable fluorophores that are stochastically activated at ultra-low densities. The bright fluorophores are sequentially localized by several MINFLUX iterations with successively decreasing $L$. Because of its higher photon efficiency, MINFLUX can improve the localization precision compared to camera-based SMLM when the resolution is photon limited. This comes at the expense of a much smaller imaged FOV and thus the recording of larger FOVs is slow compared to camera-based techniques. Therefore, precise drift compensation is especially crucial for MINFLUX imaging.

The pioneering work that introduced MINFLUX resolved fluorophore distances of 6 nm on DNA origami, achieving an average localization precision of 1.2 nm with 1000 photons[12]. In cells, MINFLUX resolved nuclear pore complex test structures[28] with a precision of under 2 nm[21,29–33], easily showing the eight-fold symmetry of NPCs (Figure 3a). These were the first structures imaged in living cells that provided a clearly superior resolution (Figure 3b) compared to previous live-cell SMLM[21,28]. On purified proteins, distances in the range of 1 – 15 nm between up to four fluorophores have been measured with a angstrom precision, demonstrating that MINFLUX can be a powerful alternative to Förster Resonance Energy Transfer (FRET)[34].

As biology is not two-dimensional, the extension of MINFLUX to 3D[21,24,29,35,36] was of high importance. In 3D MINFLUX imaging, a so-called 3D donut or "bottle beam" – effectively a sphere of light with a dark center – is positioned around the fluorophore in 3D to localize it in x, y, and z[21,29]. Equally important was the extension of MINFLUX to multiple colors for studying protein interactions and spatial context. Chromatic shifts between the color channels can be minimized by using spectrally close fluorophores that are excited with the same wavelength, and emission events are assigned to the respective fluorophore depending on the ratio obtained from spectral splitting with a dichroic mirror. Using this approach, Pape et al. investigated the proximity of different mitochondrial subcomplexes[35].

Combined with DNA-PAINT[9], where fluorophores transiently attach to the target via DNA strands, MINFLUX can record multicolor images of fixed cells by sequentially using different DNA strands for different targets. However, for such bright labels the advantage of MINFLUX compared to camera-based approaches diminishes. For example, the camera-based DNA-PAINT approach RESI[10], which sequentially images sparse target subsets at moderate spatial resolution and then averages localizations in close proximity in each of these subsets, achieved angstrom resolution on large FOVs (Figure 3c).

Commercialization of MINFLUX boosted the accessibility of the technology[29] and allowed its application to numerous biological systems including the photoreceptor active zone[36] and the injectiosome, a bacterial molecular machine[37]. Furthermore, statistical analysis of MINFLUX images of the PIEZO1 ion channel resolved few-nanometer conformational shifts caused by chemical and mechanical modulators, giving better insight into the activation of the mechanosensitive channel[38] (Figure 3d).

MINFLUX imaging shows its full potential when fluorophores of low brightness need to be used[21], when small FOVs are sufficient, and when labeling schemes with minimal linkage error can be employed. Although it has provided nice results on fixed samples, MINFLUX is hard to use for dynamic imaging in live cells because of its slow speed. Any motion of the target structure of more than the localization precision within the measurement time will lead to motion blur. However, as the measurement time scales with the FOV, tiny regions of interest, for example a single nuclear pore complex, can be imaged relatively quickly (Figure 3b). In the future, MINFLUX has the potential to outperform SMLM for dynamic live-cell imaging with a targeted activation/deactivation approach (Figure 3e): here, photo-switchable fluorophores are illuminated with a high intensity of the UV and imaging laser. After the on-switching of a fluorophore, the UV laser is stopped to avoid activation of a second fluorophore. After its localization and off-switching/bleaching with the imaging laser, the next emitter can be activated and localized in rapid succession. Such an approach necessitates new and faster microscopes, but most importantly live-cell compatible fluorophores that perform well for MINFLUX and that can be switched on rapidly by light. On the other hand, for imaging of protein assemblies in fixed cells, there might be only few applications where MINFLUX will outperform state-of-the art SMLM (Figure 3c), for example when imaging photoactivatable fluorescent proteins (PAFP), or when high 3D resolution is required.

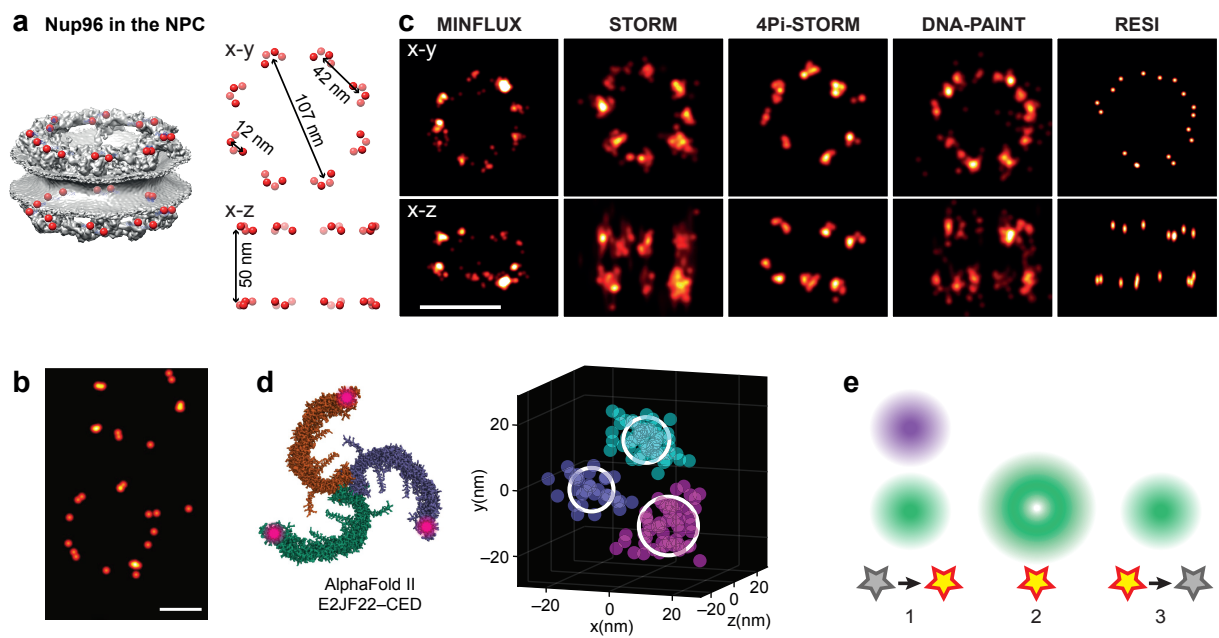

**Figure 3. MINFLUX imaging achieves molecular resolution. a,** (Left) EM density of the nuclear pore complex (NPC) with C-termini of Nup96 indicated in red. (Right) Side and top view schematic showing eightfold symmetry and distances between Nup96 proteins. **b,** NPC imaged in live cells with MINFLUX[21]. **c,** Comparison between different super-resolution techniques imaging Nup96 in NPCs in fixed cells[10,29,68,69]. **d,** (Left) Structural model of ion channel PIEZO1. Labels are shown as magenta stars. (Right) Representative PIEZO1 trimer recorded with 3D MINFLUX. **e,** Proposed targeted activation / deactivation scheme for live-cell imaging. High intensity UV activation (purple) is used with the excitation laser (green) until a fluorophore on-switching event is detected, upon which the UV laser is switched off (1). MINFLUX localization is performed until enough photons are detected (2). High-intensity illumination with the excitation laser quickly switches the fluorophore off (3) and the cycle is repeated. Scale bars: 100 nm.

## MINFLUX tracking

An application where MINFLUX already clearly outperforms camera-based localization is single-fluorophore tracking. During MINFLUX tracking, the position of the scan pattern is recentered to the estimated fluorophore position after each iteration and thus the MINFLUX pattern effectively follows the fluorophore. Compared to camera-based tracking, the high photon efficiency leads to three advantages: a high spatial precision with few detected photons, high temporal resolution because detection of fewer photons is faster, and long tracks because the fluorophore close to the dark donut center is not bleached quickly. The drawback is that ultra-low fluorophore concentrations are required.

The first MINFLUX study demonstrated tracking of single 30S ribosomal subunits labeled with a fluorescent protein in living bacteria[12]. On DNA origami test structures (Figure 4a), MINFLUX tracking achieved a localization precision of ~2 nm with a 400 μs time resolution using <100 fluorescent photons per localization[39] (Figure 4b).

Because single-color tracking cannot easily distinguish conformational changes of proteins from overall motion, it is especially useful to study either diffusion[12,29] or conformational changes when the overall motion is well defined, as for motor proteins such as kinesin (Figure 4c). Its stepping motion has been studied extensively *in vitro* on purified proteins[40–43], but many details are still debated, e.g. due to the large (bead) label size or low temporal resolution of the established techniques. Some of these controversies have been resolved by MINFLUX's high spatio-temporal resolution combined with a small fluorophore label. For example, kinesin substeps lasting only a few milliseconds were observed via MINFLUX tracking (Figure 4d) and it was concluded that this motor protein binds its fuel, ATP, during this state[44]. Furthermore, a concurrent rightward rotation of the so-called "heads" and a rotation of the stalk were identified by observing fast movements on the scale of only a few nanometers. A switching of protofilament lanes was observed on purified proteins[27] and in fixed cells[45].

Because of insufficient spatial and temporal resolution of live-cell compatible tracking techniques, kinesin stepping had not been previously observed directly in cells. Live-cell MINFLUX tracking (Figure 4e-g) overcame this limitation and resolved the 16 nm steps in 2D and in 3D[45] and allowed characterizing step size and dwell time distributions. In addition, motors were tracked in motorPAINT[46], where labeled motors walk on microtubules of fixed and permeabilized cells, and were used to map out cellular microtubules with down to protofilament resolution.

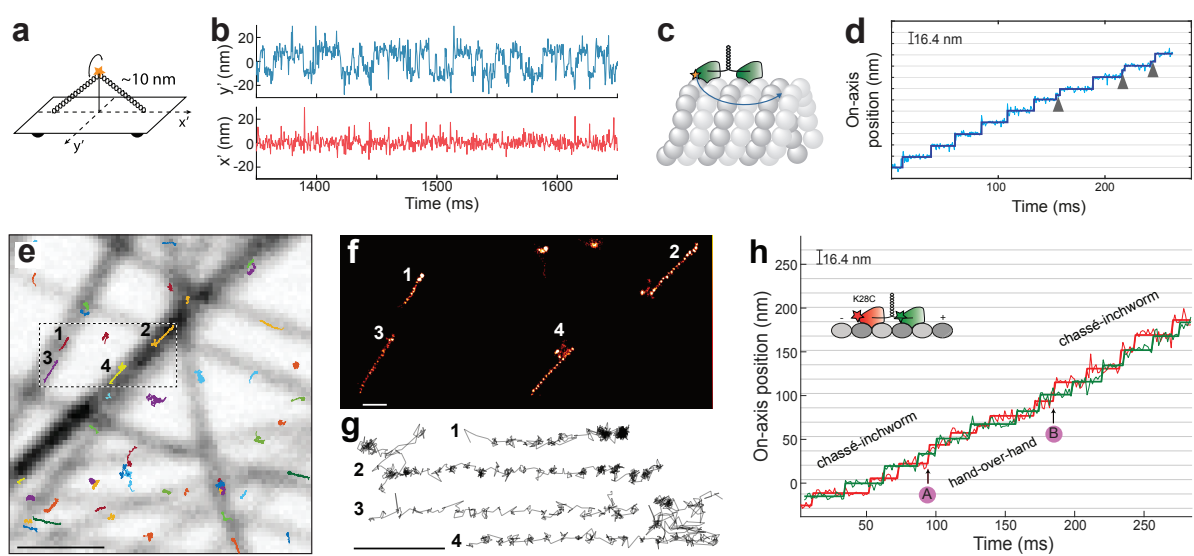

**Figure 4. MINFLUX tracking follows fast movements. a,** Illustration of the DNA origami construct with a single ATTO 647N fluorophore attached at the center of the bridge (10 nm from the origami base). The emitter can move on a half-circle above the origami and is thus restricted to a projected 1D movement. **b,** MINFLUX traces showing clear oscillation of the fluorophore on the origami along y' and negligible movement along x' with a standard deviation along the principal axis of movement of ~1.7 nm using on average 94 photons detected within 400 µs bins. **c,** Illustration of the molecular motor kinesin-1 with its two heads (green) on a microtubule (grey). **d,** Kinesin-1 trace (1 mM ATP) showing substeps (arrows). The raw data points (faint blue line) are overlaid with a 5-ms moving median filter (bold blue line). **e,** Live-cell tracks of kinesin (HaloTag-K560) show 16 nm steps. Colored tracks are overlaid with confocal images of the microtubules. Scale bar: 1 µm. **f,g,** The tracks indicated in (e) rendered as a super-resolution image (f) and line plots connecting localizations (g), showing clear walking steps (localization precision: 2 nm; temporal resolution: 1 ms). Scale bars: 100 nm. **h,** One-dimensional dual-color trace of kinesin, walking in an *in vitro* microtubule assay. The heads are labeled with different dyes and tracked simultaneously. Between Ⓐ and Ⓑ, kinesin walks in a hand-over-hand mode with the rear always overtaking the leading head. The other parts of the trace are best explained by a chassé-inchworm mechanism, where the rear lags the leading head.

Dual-color MINFLUX co-tracking[47,48] will be the next transformative development of MINFLUX, as it allows studying conformational changes of proteins during their action, ultimately in living cells. Labeling of two parts of a protein or protein complex with two different dyes and tracking them simultaneously allows for determining of their relative positions with nanometer spatial and (sub-)millisecond temporal resolution. A first implementation of MINFLUX co-tracking[47] using an in vitro assay observed conformational changes that suggest a novel walking mode of kinesin-1 (Figure 4h), highlighting the potential of MINFLUX to enable new biological insights.

## New approaches to MINFLUX

MINFLUX is a new technology, and the development of improved concepts, implementations and analysis approaches is already an active field of research. In the following we will discuss new technical implementations that will improve the performance and accessibility of MINFLUX and enable novel applications.

With a spatial resolution in the single nanometer range, MINFLUX rivals FRET in studying small distances between fluorophores, with the added advantage that the MINFLUX distance measurement is absolute and not limited in scale[34]. Pulsed interleaved MINFLUX (p-MINFLUX) allows for simultaneous MINFLUX and FRET tracking[49] (Figure 5a,b). The microscope was implemented by modifying the excitation path of a time-correlated single-photon counting confocal microscope[50] and performing fast scanning by using four beam paths of different lengths, so that the laser pulses arrive at the four scanning positions at different times. The advantage is a simple setup, but this comes at the expense of flexibility, as the scan size $L$ cannot be changed during the experiment – hence, zooming-in on the fluorophore to increase photon efficiency is not possible.

Simplifying MINFLUX microscopes will be key to enable more groups to endeavor in technical developments and to give access to the many potential biological users. An important step in this direction was the simple extension of standard confocal microscopes with minor modifications to enable MINFLUX-like measurements in a technique called RASTMIN[51], although it has to be seen if such an implementation has sufficient flexibility to unlock the full potential of MINFLUX. The development of a fast variable phase plate has the potential to enable multi-color 3D MINFLUX with a stable and simple setup[24].

Although MINFLUX is often associated with a donut-shaped PSF, it is by no means necessary and alternative PSFs featuring a local minimum might even outperform donut-based MINFLUX[24,44] (Figure 5c). Using destructive interference of two laser beams in the focus creates a line of minimum intensity that can be quickly scanned by changing the phase of one of the beams[44]. In a similar way, a bi-lobed PSF can be generated and scanned rapidly by a variable phase plate[24]. This allows for a 1D MINFLUX localization. To enable 2D and 3D MINFLUX, separate patterns for each dimension are used sequentially. 4Pi-MINFLUX[52] creates striped PSFs along three directions and comes with the advantage of doubling the detection efficiency by using two opposing objectives and hence achieving the highest 3D precision per detected photon to date (0.5 nm, 0.5 nm, 0.3 nm in x-, y- and z-directions, respectively, for ~400 photons per dimension).

Because only a single beam is used, the strongest limitations on MINFLUX are low throughput and long measuring times. Acquisitions could be drastically sped up by parallelization, i.e. simultaneous imaging with multiple beams. This is challenging, as the scan path of each beam would need to be updated in real time and centered at the corresponding fluorophore. Modulation-enhanced SMLM methods[53–55] can be thought of as a combination of structured illumination microscopy with SMLM. However, as they lack fluorophore-specific feedback, they achieve, on average, only two-fold improvements in resolution over SMLM and thus cannot be considered parallelized versions of MINFLUX.

MINFLUX needs ultra-low concentrations of fluorophores because background fluorophores in the periphery of the donut can easily be much brighter than the fluorophore of interest at the intensity minimum. Interestingly, it has been demonstrated that multiple fluorophores of the same color can be resolved with multi-emitter MINFLUX, provided that they are in close vicinity[56]. For multi-emitter fitting in SMLM, the added shot noise of both fluorophores quickly degrades the precision[57]. In contrast, keeping the fluorophores close to the minimum in MINFLUX, where one of the fluorophores can effectively be "turned off" when centered at the minimum, allows for a far better separation than in multi-emitter SMLM (Figure 5d). This approach could become a useful alternative to multi-color co-tracking, as it allows tracking distances among several positions on a single protein labeled with the same color. It needs to be seen how this approach performs under high background conditions found in living cells and how it scales with the number of emitters.

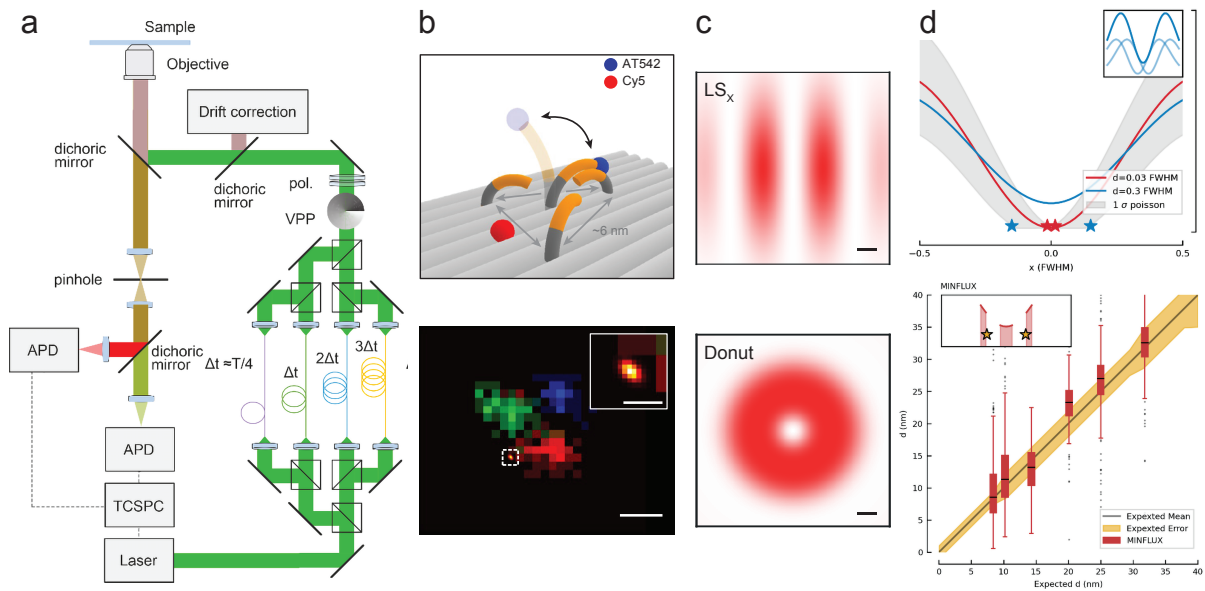

**Figure 5. New approaches to MINFLUX with simplified setups, alternative PSF shapes and multi-emitter separation. a,** Illustration of the pMINFLUX setup. A pulsed laser is split into four beams and coupled into optical fibers, which delay the laser pulses. The beams are recombined, and doughnut-shaped beams are created with a vortex phase plate (VPP). The beams are focused onto the sample in a triangular pattern with one beam in the center of the triangle. **b,** Super-resolved FRET in pMINFLUX. (Top) Illustration of the dynamic DNA origami with three protruding strands spaced ~6 nm apart, to which an ATTO 542 (AT542)-labelled DNA pointer transiently hybridizes. (Bottom) MINFLUX localizations of the DNA pointer (blue, green, red). Scale bar: 5 nm. The position of Cy5 was deduced from the calculated FRET distances (boxed inset, scale bar: 1 nm). **c,** Comparison of the line-shaped minimum $LS_x$ and donut excitation intensity distribution in the focal plane. Scale bars: 100 nm. **d,** Multi-emitter MINFLUX. (Top) Two non-blinking emitters/scatterers of the same color can be localized simultaneously below the diffraction limit as one of them can be "turned off" by the excitation minimum. The inset shows the profile of the average intensity profiles of each point scatterer and their joint signal. (Bottom) Boxplot of the measured over the expected distances between fluorophores on nano rulers of different lengths. Recording the photons originating near the illumination minimum (see inset) suffices to separate simultaneously emitting fluorophores down to 8 nm.

## Conclusion

Because of its photon efficiency in localizing single fluorophores, MINFLUX has a high potential to become a key technology for structural biology, ideally complementing electron microscopy and other super-resolution techniques. In fixed cells, it reveals precise arrangements of proteins, but optimized SMLM techniques with bright labels currently achieve a similar, if not better, resolution[10] combined with a much larger FOV. The standout feature of MINFLUX imaging might become fast live cell imaging of cellular structures in a small FOV. Currently, the most transformative impact of MINFLUX is in single-fluorophore tracking, with the development of multi-color co-tracking opening the possibility to directly watch conformational changes of proteins performing their functions in living cells.

As a young technology, MINFLUX is continuously evolving. To become a standard technique with high impact in structural biology, several challenges still need to be overcome. Live-cell MINFLUX would greatly profit from improved fluorophores for dynamic live-cell imaging and higher robustness against auto-fluorescent background and higher labeling densities. Multi-fluorophore tracking, either using separate colors or multi-emitter MINFLUX, is limited by the need for very high labeling efficiencies (so that a sufficiently high fraction of targets is labeled in all colors) in combination with ultra-low densities, which is difficult to control in living cells. Throughput is still much lower compared to camera-based techniques and new parallelization approaches, especially for imaging of larger structures, will be crucial to reduce measurement times to acceptable levels. Often, the accuracy in MINFLUX experiments is not limited by the detected photons, but by instabilities of the setup. Thus, ultra-stable microscopes will be key to reach the full potential of MINFLUX. In-depth theoretical analysis of the MINFLUX principle in combination with simulations might yield a way to improve accuracy, robustness and throughput of the technique.

Commercialization of MINFLUX has improved accessibility and ease-of-use. Further reducing complexity and difficulty of MINFLUX experiments will help to enable non-expert users to obtain data of the highest quality. The development of a robust, affordable, and easy-to-build open source MINFLUX system would further spread the use of this technology to researchers for which a commercial system is out of reach.

With these developments, the future of MINFLUX looks bright as a standard technique for structural biology that can bring molecular resolution to the living cell.

## Acknowledgements

The authors wish to thank Dr. Sheng Liu, Nikolay Sergeev, Dr. Takahiro Deguchi, and Dr. Francesco Reina for helpful discussions. This work was supported by the European Research Council (grant no. ERC CoG-724489 to J.R.).

## Competing interests

The authors declare no competing interests.

## Contributions

All authors contributed to literature research, discussion, writing and editing of the manuscript.